\documentclass{cimento}

\usepackage{graphicx}  
\usepackage{slashed}
\usepackage{amsmath}
\usepackage{amssymb}
\title{
Radiative corrections to di-meson tau decays}
\author{A.~Miranda~~\from{ins:x} }
\instlist{\inst{ins:x} 
Institut de Física d’Altes Energies (IFAE) and The Barcelona Institute of Science and
Technology (BIST), Campus UAB, 08193 Bellaterra (Barcelona), Spain.
  }

\begin{document}

\maketitle

\begin{abstract}
We review radiative corrections to tau decays into two mesons discussing their impact in new physics searches.
\end{abstract}

\section{Introduction}
Radiative corrections are needed to make new physics tests more precise and required to get the most from experimental data with below percent accuracy.

The tau is the only lepton massive enough to decay into mesons, giving thus access to CKM unitarity tests (through $V_{us}$, or $V_{us}/V_{ud}$), in addition to those of lepton universality and searches for non-standard interactions.
One-meson processes depend only on the corresponding meson decay constant, which is best determined on the lattice (without any pollution from possible new physics) \cite{ParticleDataGroup:2022pth}. Their radiative corrections have been improved recently \cite{Arroyo-Urena:2021nil,Arroyo-Urena:2021dfe} (see also \cite{Guo:2010dv, Guevara:2013wwa,Guevara:2021tpy}).

Here we will focus on the radiative corrections to the two-meson tau decays, mostly following Ref.~\cite{Escribano:2023seb} (see also \cite{Miranda:2020wdg}, only for the $\pi^-\pi^0$ case). In the corresponding non-radiative processes, dispersive form factors matching accurately precise experimental data \cite{ParticleDataGroup:2022pth} have been constructed~\cite{Jamin:2001zq,Moussallam:2007qc,Boito:2008fq,Boito:2010me,GomezDumm:2013sib,Antonelli:2013usa,Escribano:2013bca,Bernard:2013jxa,Escribano:2014joa,Descotes-Genon:2014tla,Escribano:2016ntp,Gonzalez-Solis:2019iod} and are a key input in this endeavor. These are the functions $F_{0,+}(t)$ in Eq.~(\ref{eq.Hnu}), carrying spin $0,\,1$, respectively.

\section{Amplitude}
The matrix element for the $\tau^{-}(P) \to P_{1}^{-}(p_{-}) P_{2}^{0}(p_{0}) \nu_\tau(q') \gamma(k,\epsilon)$ reads
\begin{equation}\label{Eq:Amplitude}\begin{split}
\mathcal{M}=&\frac{e G_F V^{*}_{ud}}{\sqrt{2}}\epsilon^{*}_\mu\bigg[(V^{\mu\nu}-A^{\mu\nu})\bar{u}(q')\gamma_\nu (1-\gamma^5) u(P)\\[1ex]
&\qquad+\frac{H_\nu(p_{-},p_{0})}{k^2-2k\cdot P}\bar{u}(q')\gamma^\nu (1-\gamma^5)(m_\tau +\slashed{P}-\slashed{k})\gamma^\mu u(P)\bigg]\ ,
\end{split}\end{equation}
$P_{1,2}$ being pions, kaons or eta mesons, and the hadron vector defined as
\begin{equation}\label{eq.Hnu}
H^\nu(p_{-},p_{0})=C_V F_{+}(t)Q^\nu+C_S\frac{\Delta_{{-}{0}}}{t}q^\nu F_{0}(t)\ ,
\end{equation}
with $q^\nu=(p_-+p_0)^\nu, t=q^2, Q^\nu=(p_--p_0)^\nu-\frac{\Delta_{-0}}{t}q^\nu,\Delta_{12}=m_1^2-m_2^2$, and $C_{V,S}$ is the corresponding Clebsch-Gordan coefficient (CGc), $C_{V,S}^{\pi^-\pi^0,K^-\pi^0,\bar{K}^0\pi^-,K^-K^0}=\lbrace 1,1/\sqrt{2},1,-1\rbrace$. $(V,A)_{\mu\nu}$ can be split into their model-dependent and model-independent parts, with the structure-independent one fulfilling the Low and Burnett-Kroll theorems. This one yields~\cite{Cirigliano:2002pv,Guevara:2016trs}~($t'=(P-q)^2$)
\begin{equation}\begin{split}\label{eq.VmunuSI}
V^{\mu\nu}_\text{SI}=&\frac{H^\nu(p_{-}+k,p_{0})(2p_{-}+k)^\mu}{2k\cdot p_{-}+k^2}+\left\{-C_V F_{+}(t^\prime)-\frac{\Delta_{-0}}{t^\prime}
\left[C_S F_{0}(t^\prime)-C_V F_{+}(t^\prime)\right]\right\} g^{\mu\nu}\\[1ex]
&+\frac{\Delta_{-0}}{t t^\prime}\bigg\lbrace
2\left[C_S F_{0}(t^\prime)-C_V F_{+}(t^\prime)\right]-\frac{C_S t^\prime}{k\cdot(p_{-}+p_{0})}\left[F_{0}(t^\prime)-F_{0}(t)\right]\bigg\rbrace 
q^\mu q^\nu\ \\[1ex]
&-C_V\frac{F_{+}(t^\prime)-F_{+}(t)}{k\cdot (p_{-}+p_{0})}Q^\nu q^\mu \, .\\
\end{split}\end{equation}
The (axial-)vector tensor can be decomposed in terms of four independent Lorentz structures, whose coefficients are the corresponding form factors $(\left\lbrace a_i,v_i\right\rbrace_{i=1,...,4})$ \cite{Miranda:2020wdg}. This hadron input is obtained from Chiral Perturbation Theory \cite{Weinberg:1978kz, Gasser:1983yg, Gasser:1984gg} including resonances \cite{Ecker:1988te,Ecker:1989yg}.

\section{Decay rate and spectra}\label{sec:DR&S}
In Refs.~\cite{Escribano:2023seb,Miranda:2020wdg} we predict branching ratios and spectra (in $E_\gamma$ and $t$) for different cuts on the photon energy. Although the non-radiative modes are just related by a CGc at low energies, this is no longer the case including photon corrections. Particularly, a relative sign in the $g^{\mu\nu}$ part in Eq.~(\ref{eq.VmunuSI}) between the $K^-\pi^0$ and $\bar{K}^0\pi^-$ is needed to understand that the latter can have an order of magnitude larger branching ratio than the former for $E_\gamma^{cut}\geq 300$ MeV \cite{Escribano:2023seb}.

We emphasize the importance of any measurement of these decays to reduce the uncertainty on the radiative corrections and therefore increase the reach on new physics searches (as \cite{Miranda:2020wdg,Masjuan:2023qsp} exemplify in the case of the muon $g-2$ when using the di-pion tau decay for the corresponding isospin-rotated contribution to its hadronic vacuum polarization piece).

\section{Radiative corrections}
The $G_{\text{EM}}(t)$ function encapsulating the long-distance electromagnetic corrections (from on- and off-shell photons) is defined in the usual way \cite{Cirigliano:2001er}

\begin{equation}\begin{split}
\label{eqGEM}
\frac{d\Gamma}{dt}
\Bigg|_{PP(\gamma)}&=
\frac{G_{F}^2\left\vert V_{uD}F_{+}(0)\right\vert^2 S_{\rm EW}m_\tau^3}{768\pi^3 t^3}
\left(1-\frac{t}{m_\tau^2}\right)^2\lambda^{1/2}(t,m_{-}^2,m_{0}^2)\\[1ex]
&\quad\times\left[C_V^2\vert\tilde{F}_{+}(t)\vert^2\left(1+\frac{2t}{m_\tau^2}\right)
\lambda(t,m_{-}^2,m_{0}^2)+3 C_S^2\Delta_{-0}^2\vert\tilde{F}_{0}(t)\vert^2\right] G_{\rm EM}(t)\ ,
\end{split}\end{equation}
where $S_{EW}$ encodes the universal short-distance electroweak corrections \cite{Marciano:1988vm,Erler:2002mv}, $V_{uD}\,(D=d,s)$ is the appropriate CKM matrix element and $G_F$ is the Fermi constant. We also divide $G_{\text{EM}}(t)$ into a part including the non-radiative plus the virtual contributions together with the leading Low approximation, $G_{\text{EM}}^{(0)}(t)$, and $\delta G_{\text{EM}}(t)$, which has all remaining pieces.\\

\begin{figure}[h!]
\centering			
\includegraphics[width=0.47\linewidth]{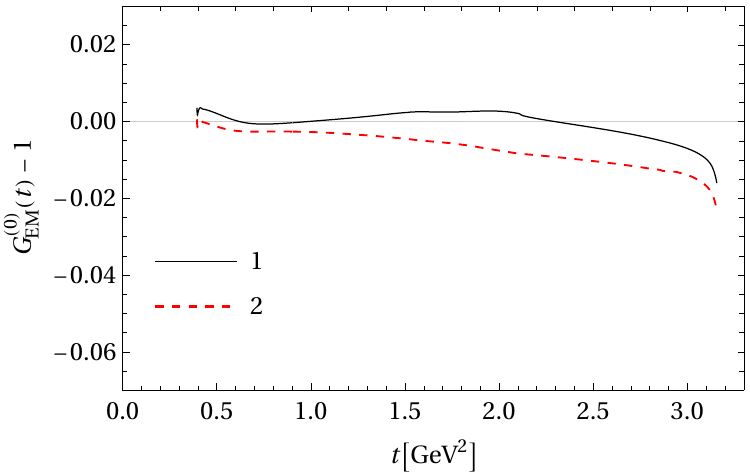}\qquad
\includegraphics[width=0.47\linewidth]{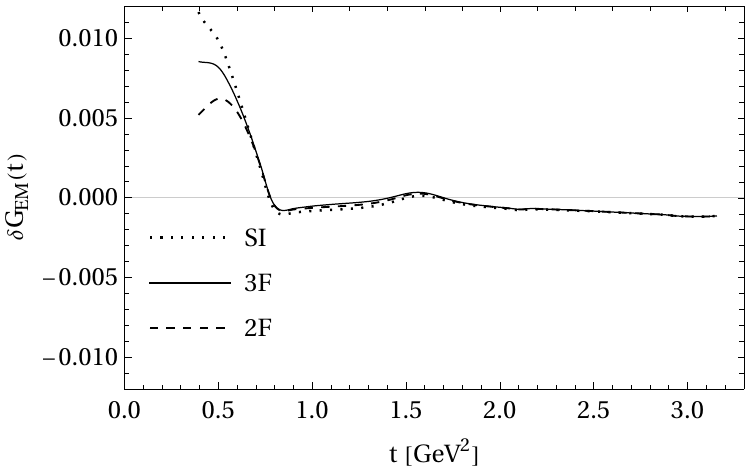}\\[1ex]	
\includegraphics[width=0.47\linewidth]{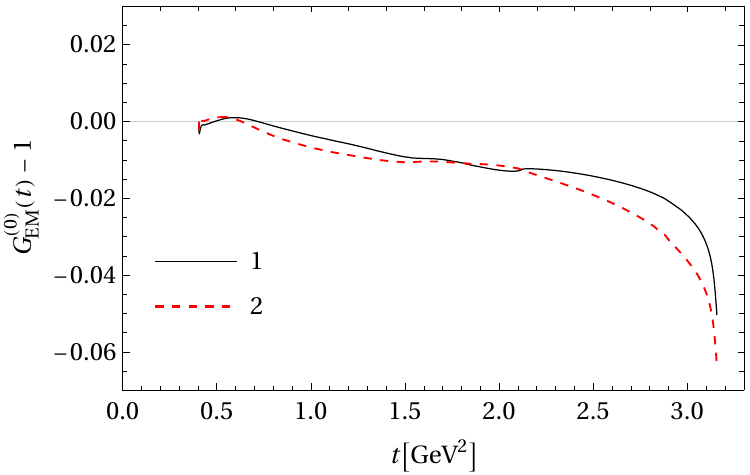}\qquad
\includegraphics[width=0.47\linewidth]{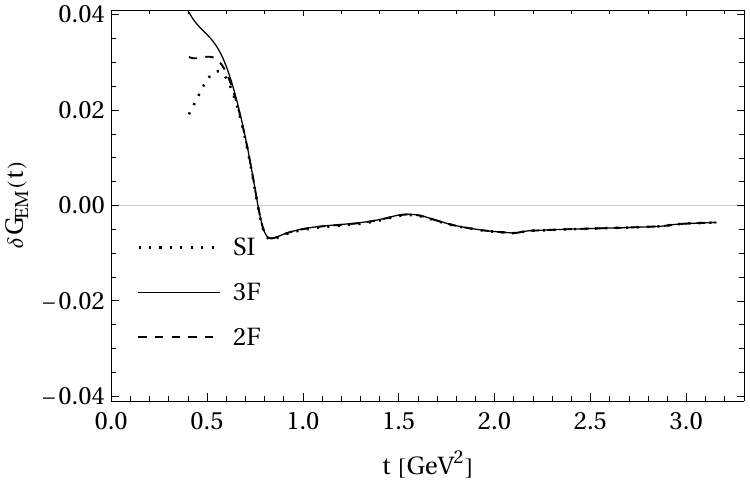}
\\
\caption{Correction factors $G_{\rm EM}^{(0)}(t)$ (left) and $\delta G_{\rm EM}(t)$ (right) 
to the differential decay rates of the $K^{-}\pi^0$ (top) and $\bar{K}^{0}\pi^{-}$ (bottom) modes.}
\label{fig:RADCOR}
\end{figure}

In figure \ref{fig:RADCOR}, we plot the $G_{\text{EM}}^{(0)}(t)$ function and its non-leading part, $\delta G_{\text{EM}}$, for the $K\pi$ tau decay modes. The corresponding plots for the di-pion and di-kaon channels can be found in Refs.~\cite{Escribano:2023seb,Miranda:2020wdg}, respectively. The sizable difference between the results for both $K\pi$ modes has the same explanation given in section \ref{sec:DR&S}.
\\

We define the radiative correction $\delta_{\rm EM}^{PP}$ by integrating upon the di-meson invariant mass, to get
\begin{equation}
\Gamma_{PP(\gamma)}=
\frac{G_F^2 S_{\rm EW}m_\tau^5}{96\pi^3}\left\vert V_{uD} F_{+}(0)\right\vert^2 I^{\tau}_{PP}
\left(1+\delta_{\rm EM}^{PP}\right)^2\ ,
\end{equation}
where the analytical expression for the $I^\tau_{PP}$ integral can be found in \cite{Escribano:2023seb}.

Our main outcomes are \begin{eqnarray}
\label{eq:MAINRESULT}
&\delta_{\rm EM}^{K^-\pi^0}=-\left(0.009^{+0.008}_{-0.118}\right)\%\ ,
\qquad\delta_{\rm EM}^{\bar{K}^0\pi^-}=-\left(0.166^{+0.010}_{-0.122}\right)\%\ ,\nonumber\\[1ex]
&\delta_{\rm EM}^{K^-K^0}=-\left(0.030^{+0.026}_{-0.179}\right)\%\ ,
\qquad\delta_{\rm EM}^{\pi^-\pi^0}=-\left(0.186^{+0.024}_{-0.169}\right)\%\ ,
\end{eqnarray}
which agree with previous results \cite{Antonelli:2013usa,Flores-Baez:2006yiq}, halving the uncertainty for the $K\pi$ cases (ours is the first evaluation for the $K^-K^0$ channel). $\delta^{K^-\eta^{(')}}$ is also estimated in \cite{Escribano:2023seb}, based on the dominant contribution in each channel. Because electromagnetic corrections are of the same order as $G$-parity breaking effects, $\delta^{\pi^-\eta^{(')}}$ can only be computed after the leading photonic contributions have been subtracted \cite{Guevara:2016trs}, which goes beyond our scope, as it would then be a permille effect on the still unmeasured $\tau^-\to\pi^-\eta^{(')}\nu_\tau$ decays.

\section{Impact on new physics searches}
$\tau^-\to\bar{u}d\nu_\tau$ decays can be studied in a generalized Fermi-type theory including the most general interactions of (pseudo)scalar, (axial-)vector and antisymmetric tensor type \cite{Cirigliano:2009wk,Garces:2017jpz,Cirigliano:2017tqn,Miranda:2018cpf,Cirigliano:2018dyk,Rendon:2019awg,Chen:2019vbr,Gonzalez-Solis:2019lze,Gonzalez-Solis:2020jlh,Chen:2020uxi,Chen:2021udz,Cirigliano:2021yto,Arteaga:2022xxy}. Doing this, the new physics scale is restricted (under the weak-coupling assumption) to be $\Lambda\gtrsim[2,4]$ TeV, depending on the particular channel(s) and observables analyzed, at $90\%$ C.L., and radiative corrections do not yet affect these limits significantly \cite{Arroyo-Urena:2021nil,Arroyo-Urena:2021dfe,Escribano:2023seb}. However, they increase the overall agreement with the SM, mostly through their impact on scalar non-standard interactions. The implications of the radiative corrections to di-meson tau decays in lepton universality \cite{Arroyo-Urena:2021nil,Arroyo-Urena:2021dfe} and CKM unitarity tests \cite{Antonelli:2013usa} remain to be studied.
\section{Conclusions}
We have first computed the complete radiative corrections to the $\tau^-\to (P P')^-\nu_\tau$ decays, including the structure-dependent corrections, complying with the constraints from chiral symmetry at low $t$ and with Low's and Burnett-Kroll theorems at low $E_\gamma$. Our main results are Eqs.~(\ref{eq:MAINRESULT}), complementing previous evaluations for the one-meson modes by our group \cite{Arroyo-Urena:2021nil,Arroyo-Urena:2021dfe}, enabling more precise new physics searches.
\acknowledgments
I wish to thank the organizers for the pleasant conference. I
also would like to acknowledge R. Escribano and P. Roig for their collaboration and 
comments on the manuscript. This work has been supported by the Ministerio de
Ciencia e Innovación under the grant PID2020-112965GB-
I00, and by the European Union’s Horizon 2020 Research and
Innovation Programme under the grant no. 824093 (H2020-
INFRAIA-2018-1). IFAE is partially funded by the CERCA 
program of the Generalitat de Catalunya.

A.M. acknowledges the support from the Departament de Recerca i Universitats de la Generalitat de Catalunya al Grup de Recerca i Universitats from Generalitat de Catalunya to the Grup de Recerca 00649 (Codi: 2021 SGR 00649).


\begin{thebibliography}{0}
\bibitem{ParticleDataGroup:2022pth}
R.~L.~Workman \textit{et al.} [Particle Data Group],
PTEP \textbf{2022} (2022), 083C01.

\bibitem{Arroyo-Urena:2021nil}
M.~A.~Arroyo-Ure\~na, G.~Hern\'andez-Tom\'e, G.~L\'opez-Castro, P.~Roig and I.~Rosell,
Phys. Rev. D \textbf{104} (2021) no.9, L091502.

\bibitem{Arroyo-Urena:2021dfe}
M.~A.~Arroyo-Ure\~na, G.~Hern\'andez-Tom\'e, G.~L\'opez-Castro, P.~Roig and I.~Rosell,
JHEP \textbf{02} (2022), 173.

\bibitem{Guo:2010dv}
Z.~H.~Guo and P.~Roig,
Phys. Rev. D \textbf{82} (2010), 113016.

\bibitem{Guevara:2013wwa}
A.~Guevara, G.~L\'opez Castro and P.~Roig,
Phys. Rev. D \textbf{88} (2013) no.3, 033007.

\bibitem{Guevara:2021tpy}
A.~Guevara, G.~L.~Castro and P.~Roig,
Phys. Rev. D \textbf{105} (2022) no.7, 076007.

\bibitem{Escribano:2023seb}
R.~Escribano, A.~Miranda and P.~Roig,
[arXiv:2303.01362 [hep-ph]].

\bibitem{Miranda:2020wdg}
J.~A.~Miranda and P.~Roig,
Phys. Rev. D \textbf{102} (2020), 114017.

\bibitem{Jamin:2001zq}
M.~Jamin, J.~A.~Oller and A.~Pich,
Nucl. Phys. B \textbf{622} (2002), 279-308.

\bibitem{Moussallam:2007qc}
B.~Moussallam,
Eur. Phys. J. C \textbf{53} (2008), 401-412.

\bibitem{Boito:2008fq}
D.~R.~Boito, R.~Escribano and M.~Jamin,
Eur. Phys. J. C \textbf{59} (2009), 821-829.

\bibitem{Boito:2010me}
D.~R.~Boito, R.~Escribano and M.~Jamin,
JHEP \textbf{09} (2010), 031.

\bibitem{GomezDumm:2013sib}
D.~G\'omez Dumm and P.~Roig,
Eur. Phys. J. C \textbf{73} (2013) no.8, 2528.

\bibitem{Antonelli:2013usa}
M.~Antonelli, V.~Cirigliano, A.~Lusiani and E.~Passemar,
JHEP \textbf{10} (2013), 070.

\bibitem{Escribano:2013bca}
R.~Escribano, S.~Gonzalez-Solis and P.~Roig,
JHEP \textbf{10} (2013), 039.

\bibitem{Bernard:2013jxa}
V.~Bernard,
JHEP \textbf{06} (2014), 082.

\bibitem{Escribano:2014joa}
R.~Escribano, S.~Gonz\'alez-Sol\'\i{}s, M.~Jamin and P.~Roig,
JHEP \textbf{09} (2014), 042.

\bibitem{Descotes-Genon:2014tla}
S.~Descotes-Genon and B.~Moussallam,
Eur. Phys. J. C \textbf{74} (2014), 2946.

\bibitem{Escribano:2016ntp}
R.~Escribano, S.~Gonz\`alez-Sol\'is and P.~Roig,
Phys. Rev. D \textbf{94} (2016) no.3, 034008.

\bibitem{Gonzalez-Solis:2019iod}
S.~Gonz\`alez-Sol\'\i{}s and P.~Roig,
Eur. Phys. J. C \textbf{79} (2019) no.5, 436.

\bibitem{Cirigliano:2002pv}
V.~Cirigliano, G.~Ecker and H.~Neufeld,
JHEP \textbf{08} (2002), 002.

\bibitem{Guevara:2016trs}
A.~Guevara, G.~L\'opez-Castro and P.~Roig,
Phys. Rev. D \textbf{95} (2017) no.5, 054015.

\bibitem{Weinberg:1978kz}
S.~Weinberg,
Physica A \textbf{96} (1979) no.1-2, 327-340.

\bibitem{Gasser:1983yg}
J.~Gasser and H.~Leutwyler,
Annals Phys. \textbf{158} (1984), 142.

\bibitem{Gasser:1984gg}
J.~Gasser and H.~Leutwyler,
Nucl. Phys. B \textbf{250} (1985), 465-516.

\bibitem{Ecker:1988te}
G.~Ecker, J.~Gasser, A.~Pich and E.~de Rafael,
Nucl. Phys. B \textbf{321} (1989), 311-342.

\bibitem{Ecker:1989yg}
G.~Ecker, J.~Gasser, H.~Leutwyler, A.~Pich and E.~de Rafael,
Phys. Lett. B \textbf{223} (1989), 425-432
doi:10.1016/0370-2693(89)91627-4

\bibitem{Masjuan:2023qsp}
P.~Masjuan, A.~Miranda and P.~Roig,
[arXiv:2305.20005 [hep-ph]].

\bibitem{Cirigliano:2001er}
V.~Cirigliano, G.~Ecker and H.~Neufeld,
Phys. Lett. B \textbf{513} (2001), 361-370.

\bibitem{Marciano:1988vm}
W.~J.~Marciano and A.~Sirlin,
Phys. Rev. Lett. \textbf{61} (1988), 1815-1818.

\bibitem{Erler:2002mv}
J.~Erler,
Rev. Mex. Fis. \textbf{50} (2004), 200-202.

\bibitem{Flores-Baez:2006yiq}
F.~Flores-Baez, A.~Flores-Tlalpa, G.~Lopez Castro and G.~Toledo Sanchez,
Phys. Rev. D \textbf{74} (2006), 071301.

\bibitem{Cirigliano:2009wk}
V.~Cirigliano, J.~Jenkins and M.~Gonz\'alez-Alonso,
Nucl. Phys. B \textbf{830} (2010), 95-115.

\bibitem{Garces:2017jpz}
E.~A.~Garc\'es, M.~Hern\'andez Villanueva, G.~L\'opez Castro and P.~Roig,
JHEP \textbf{12} (2017), 027

\bibitem{Cirigliano:2017tqn}
V.~Cirigliano, A.~Crivellin and M.~Hoferichter,
Phys. Rev. Lett. \textbf{120} (2018) no.14, 141803.

\bibitem{Miranda:2018cpf}
J.~A.~Miranda and P.~Roig,
JHEP \textbf{11} (2018), 038.

\bibitem{Cirigliano:2018dyk}
V.~Cirigliano, A.~Falkowski, M.~Gonz\'alez-Alonso and A.~Rodr\'\i{}guez-S\'anchez,
Phys. Rev. Lett. \textbf{122} (2019) no.22, 221801.

\bibitem{Rendon:2019awg}
J.~Rend\'on, P.~Roig and G.~Toledo S\'anchez,
Phys. Rev. D \textbf{99} (2019) no.9, 093005.

\bibitem{Chen:2019vbr}
F.~Z.~Chen, X.~Q.~Li, Y.~D.~Yang and X.~Zhang,
Phys. Rev. D \textbf{100} (2019) no.11, 113006.

\bibitem{Gonzalez-Solis:2019lze}
S.~Gonz\`alez-Sol\'\i{}s, A.~Miranda, J.~Rend\'on and P.~Roig,
Phys. Rev. D \textbf{101} (2020) no.3, 034010.

\bibitem{Gonzalez-Solis:2020jlh}
S.~Gonz\`alez-Sol\'\i{}s, A.~Miranda, J.~Rend\'on and P.~Roig,
Phys. Lett. B \textbf{804} (2020), 135371.

\bibitem{Chen:2020uxi}
F.~Z.~Chen, X.~Q.~Li and Y.~D.~Yang,
JHEP \textbf{05} (2020), 151.

\bibitem{Chen:2021udz}
F.~Z.~Chen, X.~Q.~Li, S.~C.~Peng, Y.~D.~Yang and H.~H.~Zhang,
JHEP \textbf{01} (2022), 108.

\bibitem{Cirigliano:2021yto}
V.~Cirigliano, D.~D\'\i{}az-Calder\'on, A.~Falkowski, M.~Gonz\'alez-Alonso and A.~Rodr\'\i{}guez-S\'anchez,
JHEP \textbf{04} (2022), 152.

\bibitem{Arteaga:2022xxy}
S.~Arteaga, L.~Y.~Dai, A.~Guevara and P.~Roig,
Phys. Rev. D \textbf{106} (2022) no.9, 096016.


\end{thebibliography}
\end{document}